\documentclass[12pt]{article}

\usepackage{amstext, amssymb, amsthm, amsmath,bm,graphicx,rotating,array,natbib,stmaryrd,booktabs}
\usepackage[lined,boxed,linesnumbered]{algorithm2e}

\usepackage[margin=1in]{geometry}

\newtheorem{proposition}{{\bf Proposition}}
\newtheorem{lemma}{{\bf Lemma}}
\newtheorem{corollary}{{\bf Corollary}}

\newtheorem{theorem}{{\bf Theorem}}

\newcommand{\vect}{\mathrm{vec}}

\newcommand{\real}[1]{\mathrm{I \! R} \mathit{^{#1}}}
\newcommand{\trans}{^{\mbox{\tiny {\sf T}}}}

\newcommand{\Abf}{{\bm A}}
\newcommand{\Bbf}{{\bm B}}
\newcommand{\Dbf}{{\bm D}}

\newcommand{\Ibf}{{\bm I}}

\newcommand{\Kbf}{{\bm K}}

\newcommand{\Pbf}{{\bm P}}
\newcommand{\Qbf}{{\bm Q}}
\newcommand{\Sbf}{{\bm S}}
\newcommand{\Ubf}{{\bm U}}
\newcommand{\Vbf}{{\bm V}}

\newcommand{\Xbf}{{\bm X}}
\newcommand{\Ybf}{{\bm Y}}
\newcommand{\Zbf}{{\bm Z}}

\newcommand{\abf}{{\bm a}}
\newcommand{\bbf}{{\bm b}}
\newcommand{\cbf}{{\bm c}}

\newcommand{\ubf}{{\bm u}}
\newcommand{\vbf}{{\bm v}}
\newcommand{\wbf}{{\bm w}}
\newcommand{\xbf}{{\bm x}}
\newcommand{\ybf}{{\bm y}}
\newcommand{\zbf}{{\bm z}}

\newcommand{\zerobf}{{\mathbf 0}}

\newcommand{\greekbold}[1]{\mbox{\boldmath $#1$}}

\newcommand{\gammabf}{\greekbold{\gamma}}

\newcommand{\sigmabf}{\greekbold{\sigma}}

\newcommand{\Sigmabf}{\greekbold{\Sigma}}

\newcommand{\Xibf}{\greekbold{\Xi}}

\title{\Large{\textbf{Regularized Matrix Regression}}}
\author{
Hua Zhou \\
Department of Statistics \\
North Carolina State University \\
Raleigh, NC 27695-8203 \\
{\tt hau\_zhou@ncsu.edu}
\and
Lexin Li\\
Department of Statistics \\
North Carolina State University \\
Raleigh, NC 27695-8203 \\
{\tt li@stat.ncsu.edu}
}
\date{}

\begin{document}
\maketitle

%\begin{footnotetext}[1]
%{\textit{Address for correspondence: Hua Zhou, Department of Statistics, North Carolina State University, Box 8203, Raleigh, NC 27695-8203. Email: hua\_zhou@ncsu.edu.}}
%\end{footnotetext}

\baselineskip=20pt

\begin{abstract}
Modern technologies are producing a wealth of data with complex structures. For instance, in two-dimensional digital imaging, flow cytometry, and electroencephalography, matrix type covariates frequently arise when measurements are obtained for each combination of two underlying variables. To address scientific questions arising from those data, new regression methods that take matrices as covariates are needed, and sparsity or other forms of regularization are crucial due to the ultrahigh dimensionality and complex structure of the matrix data. The popular lasso and related regularization methods hinge upon the sparsity of the true signal in terms of the number of its nonzero coefficients. However, for the matrix data, the true signal is often of, or can be well approximated by, a low rank structure. As such, the sparsity is frequently in the form of low rank of the matrix parameters, which may seriously violate the assumption of the classical lasso. In this article, we propose a class of regularized matrix regression methods based on spectral regularization. Highly efficient and scalable estimation algorithm is developed, and a degrees of freedom formula is derived to facilitate model selection along the regularization path. Superior performance of the proposed method is demonstrated on both synthetic and real examples.
\end{abstract}


\section{Introduction}

Modern scientific applications are frequently producing data sets where the sampling unit is not in the form of a vector but instead a matrix. Examples include two-dimensional digital imaging data, which contain the quantized brightness value of a given color at a number of rows and columns of pixels; and flow cytometric data, which consist of the fluorescence intensity of multiple cells at multiple channels. Our motivating example is a study of an electroencephalography (EEG) data of alcoholism (\textsf{http://kdd.ics.uci.edu/datasets/eeg/eeg.data.html}). The study consists of 122 subjects with two groups: an alcoholic group and a normal control group, and each subject was exposed to a stimulus. Voltage values were measured from 64 channels of electrodes placed on the subject's scalp for 256 time points, so each sampling unit is a $256 \times 64$ matrix. It is of scientific interest to study the association between alcoholism and the pattern of voltage over times and channels \citep{LiKimAltman2010}. The generalized linear model (GLM) \citep{McCullaghNelder83GLMBook} offers a useful tool for that purpose, where the response $Y$ is the binary indicator of alcoholic or control, and the predictors include the matrix-valued EEG data $\Xbf$ and possible covariate vector $\Zbf$ such as age and gender. However, the classical GLM deals with a vector of covariates, and the presence of matrix type covariates poses fresh challenges to statistical analysis. First, na\"{i}vely turning a matrix into a vector results in an exceedingly large dimensionality; for instance, for the EEG data, the dimension is $p = 256 \times 64 = 16,384$, whereas the sample size is only $n = 122$. Second, vectorization destroys the wealth of structural information inherently possessed in the matrix data; e.g., it is commonly expected that the voltage values of the adjacent time points and channels are highly correlated. Given the ultrahigh dimensionality and the complex structure, \emph{regularization} becomes crucial for the analysis of such data. In this article, we propose a novel regularization solution for regression with matrix covariates that efficiently tackles the ultrahigh dimensionality while preserving the matrix structure.

A large variety of regularization methods have been developed in recent years. Among them, penalization has been playing a powerful role in stabilizing the estimates,  improving the risk property, and increasing the generalization power in classical regressions. Popular penalization techniques include lasso \citep{Tibshirani96Lasso,Donoho94Wavelet}, SCAD \citep{FanLi01SCAD}, fused lasso \citep{Tibshirani05FusedLasso}, elastic net \citep{ZouHastie05Enet}, and many others. For regressions with matrix covariates, a direct approach is to first vectorize the covariates then apply the classical penalization techniques. Regularization helps alleviate the problem that the dimensionality far exceeds the sample size. However, this is unsatisfactory, since it does not incorporate the matrix structural information. More importantly, the solution is based upon a fundamental assumption that the true underlying signal is sparse in terms of the $\ell_0$ norm of the regression parameters. In matrix regressions, however, it is often the case that the true signal is of a low rank structure, or can be well approximated by a low rank structure. As such, sparsity is in terms of the \emph{rank} of the matrix parameters, which is intrinsically different from sparsity in the number of nonzero entries.

\begin{figure}[t]
\centering
\includegraphics[width=4.5in]{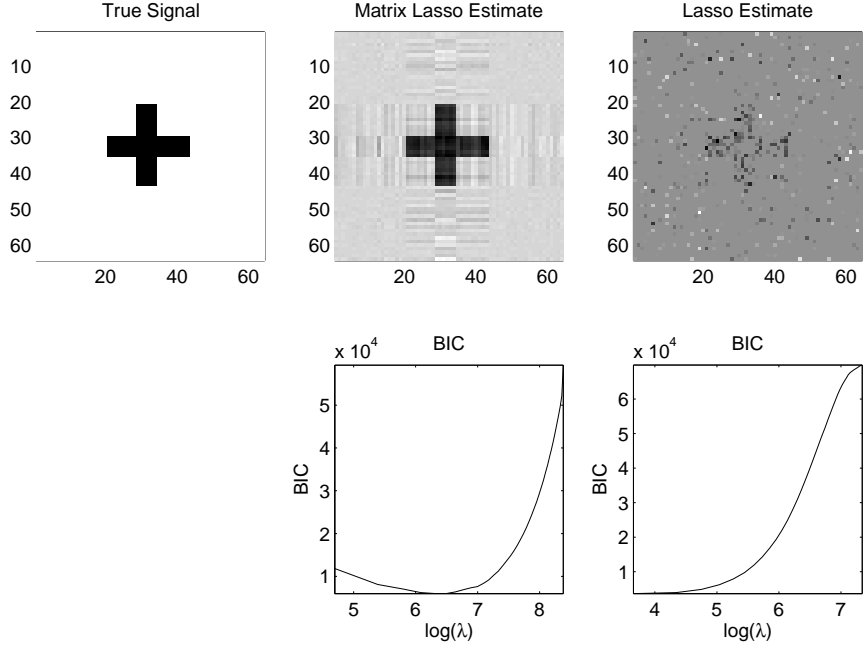}
\caption{Comparison of the nuclear norm regularized estimation to the classical lasso. The matrix covariate is $64 \times 64$, and the sample size is $500$. Top panel: True signal (left), nuclear norm regularized estimate with minimal BIC (center), and the classical lasso estimate with minimal BIC (right). Bottom panel: BIC along solution paths for the nuclear norm regularization (left) and the classical lasso regularization (right).}
\label{fig:cross-bic}
\end{figure}

To see how such a difference affects signal estimation in matrix regressions, we consider the following illustrative example. We generated a normal response $Y$ with mean, $\mu = \gammabf \trans \Zbf + \langle \Bbf, \Xbf \rangle$, and variance one. $\Zbf \in \real{5}$ denotes a usual vector of covariates with standard normal entries, and $\gammabf = (1, \ldots, 1)\trans$. $\Xbf \in \real{64 \times 64}$ denotes the matrix covariates, of which all entries are standard normal, and $\Bbf$ is the coefficient matrix of the same size. $\Bbf$ is binary, with the true signal region, which is a cross shape in our example, equal to one and the rest zero. The inner product between two matrices is defined as $\langle \Bbf,\Xbf \rangle = \langle \vect \Bbf, \vect \Xbf \rangle = \sum_{i_1,i_2} \beta_{i_1i_2} x_{i_1 i_2}$, where $\vect(\cdot)$ is the vectorization operator that stacks the columns of a matrix into a vector. We sampled $n=500$ instances $\{(y_i, \xbf_i, \zbf_i), i = 1, \ldots, 500\}$, and our goal is to identify $\Bbf$ through a regression of $y_i$ on $(\xbf_i, \zbf_i)$.  We note that our problem differs from the usual edge detection or object recognition in imaging processing \citep{Qiu2005book,Qiu2007jumpsurface}. In our setup, all elements of the image $\Xbf$ follow the same distribution. The signal region is defined through the coefficient image $\Bbf$ and needs to be inferred from the association between $Y$ and $\Xbf$ after adjusting for $\Zbf$. We also note that the total number of entries in $\Bbf$ is $4096=64^2$, and the number of nonzero ones is 240 (about $5.8\%$). We applied two approaches. The first is lasso to the vectorized $\Xbf$. That is, we solve the optimization problem
\begin{eqnarray*}
\min_{\Bbf} \, \frac 12 \sum_{i=1}^{n} (y_i - \gammabf \trans \zbf_i - \langle \Bbf, \xbf_i \rangle)^2 + \lambda \|\mathrm{vec} \Bbf \|_1,
\end{eqnarray*}
where $\|\mathrm{vec} \Bbf \|_1$ is the $\ell_1$ norm of the vectorized $\Bbf$, and $\lambda$ is the regularization parameter. The lower right panel of Figure \ref{fig:cross-bic} displays the Bayesian information criterion (BIC) along the lasso solution path, which suggests a model with the maximum number of predictors (500) that is allowed given the sample size. The parameter estimate under this model is shown in the upper right panel, which appears far away from the truth. The second solution we consider is penalizing the nuclear norm of $\Bbf$, i.e., we solve
\begin{eqnarray*}
\min_{\Bbf} \, \frac 12 \sum_{i=1}^{n} (y_i - \gammabf \trans \zbf_i - \langle \Bbf, \xbf_i \rangle)^2 + \lambda \|\Bbf\|_*,
\end{eqnarray*}
where the nuclear norm $\|\Bbf\|_* = \sum_j \sigma_j(\Bbf)$, and $\sigma_j(\Bbf)$'s are the singular values of the matrix $\Bbf$. The nuclear norm $\|\Bbf\|_*$ is a suitable measure of the ``size" of a matrix parameter, and is a convex relaxation of $\text{rank}(\Bbf) = \|\sigma(\Bbf)\|_0$. This is analogous to the $\ell_1$ norm for a vector \citep{RechtFazelParrilo10NuclearNorm}. The lower left graph of Figure \ref{fig:cross-bic} displays the BIC along the solution path of the nuclear norm penalized matrix regression, and the upper middle panel shows the corresponding estimate with minimal BIC. It is clearly seen that the nuclear norm estimate achieves a substantially better recovery than the lasso estimate. One might argue that fused lasso \citep{Tibshirani05FusedLasso} might give a better recovery of such piecewise constant signals. However, there are numerous low rank signals, e.g., $(0 1 \ldots 0 1) \trans (1 0 \ldots 1 0)$, which are extremely non-smooth and would fail fused lasso.

More generally, in this article, we propose a family regularized regression models with matrix covariates based on spectral regularization. Our contributions are multifold.  First, we employ a spectral regularization formulation and integrate within a generalized linear model (GLM) framework. The resulting model works for a variety of penalization functions, including lasso, elastic net, SCAD, and many others, as well as different types of response variables, including normal, binary and count outcomes. Second, we develop a highly efficient and scalable Nesterov algorithm for model estimation with explicit, non-asymptotic convergence rate. We emphasize that such a highly scalable algorithm is critical for analyzing large-scale and ultrahigh-dimensional matrix data. Third, we derive the effective degrees of freedom of selected models, which is crucial for tuning of the regularization parameter. The result can be viewed as an extension of the degrees of freedom development from the classical lasso model \citep{ZouHastieTibshirani07LassoDF} and the group lasso model \citep{YuanLin06GroupLasso} to the generalized linear matrix model. On the other hand, our proposal is \emph{not} simply another variant of lasso and alike. We aim at \emph{matrix regression} problems, which are important in imaging and other scientific applications but have received relatively little attention.

Our proposal is related to but also distinct from two recent developments involving matrix data. The first is a recent proposal of a family of generalized linear models with matrix or tensor (multidimensional arrays) covariates \citep{ZhouLiZhu12CPTensor}. The basic idea is to impose a particular low rank structure (CP decomposition) on $\Bbf$ and then introduce a sparse penalty on the coefficients of $\Bbf$. That solution fits the model at a \emph{fixed} rank of the matrix/tensor regression parameters, and thus corresponds to the hard thresholding in the classical vector covariate case. In contrast, our solution does not fix the rank of the parameters and is a soft thresholding procedure. Moreover, even when using a convex penalty function such as the lasso penalty, the approach of \citet{ZhouLiZhu12CPTensor} involves a challenging non-convex optimization task, whereas the solution in this paper remains a convex problem. We also note that \citet{HungWang11} considered matrix logistic regression, which is a special case of \citet{ZhouLiZhu12CPTensor}, and they did not investigate any sparsity regularization. The second related work is the line of research in matrix completion, where nuclear norm type regularization has been widely employed \citep{CandesRecht09MatrixCompletion,Mazumder09SVDReg,CaiCandesShen10SVT}. However, the two approaches are different in that, the matrix completion problem aims to recover a low rank matrix when only a small portion of its entries are observed., whereas our approach concerns about regressions with matrix covariates.

The rest of the article is organized as follows. We formulate the spectral regularization for matrix regression in Section \ref{sec:formulation}, and develop a highly scalable algorithm for the associated optimization in Section \ref{sec:algorithm}. We derive the degrees of freedom formula in Section \ref{sec:dof}, and investigate the numerical performance of the proposed method in Section \ref{sec:numerics}. We conclude the paper with a discussion of potential future research in Section \ref{sec:discussion}. We delegate all the technical proofs to the Appendix.

\section{Spectral Regularization}
\label{sec:formulation}

We first fix the notations. For any matrix $\Bbf \in \real{p_1 \times p_2}$, $\sigma(\Bbf) = (\sigma_1(\Bbf), \ldots, \sigma_q(\Bbf))$, $q = \min\{p_1,p_2\}$, denotes the vector of decreasingly ordered singular value mapping of $\Bbf$. That is $\sigma_1(\Bbf) \ge \sigma_2(\Bbf) \ge \ldots \ge \sigma_r(\Bbf) > \sigma_{r+1}(\Bbf) = \ldots = \sigma_q(\Bbf) = 0$, where $r = \mbox{rank}(\Bbf)$. %The nuclear norm is then defined as $\|\Bbf\|_* = \sum_i \sigma_i(\Bbf) = \|\sigma(\Bbf)\|_1$.
Let $Y$ denote the response variable, $\Zbf \in \real{p_0}$ the vector covariate, $\Xbf \in \real{p_1 \times p_2}$ the 2D matrix covariate, and $(y, \xbf, \zbf)$ their sample instances.

We consider the generalized linear model setup, where $Y$ belongs to an exponential family with probability mass function or density
\begin{eqnarray*}
p(y | \xbf, \zbf) = \exp\left\{ \frac{y\theta - b(\theta)}{a(\phi)} + c(y,\phi) \right\},
\end{eqnarray*}
and the first conditional moment is $\mathrm{E}(Y | \Xbf, \Zbf) = \mu = b'(\theta)$, and $\mu$ is of the form
\begin{eqnarray}
\label{eqn:tensor-reg-eta}
g(\mu) = \eta = \gammabf \trans \Zbf + \langle \Bbf,\Xbf \rangle,
\end{eqnarray}
where $g$ is a known link function, $\gammabf \in \real{p_0}$, and $\Bbf \in \real{p_1 \times p_2}$. For simplicity, in our subsequent development of the regularized matrix model estimation, we drop the vector covariate $\Zbf$ and its associated parameter $\gammabf$. However, the results can be extended straightforwardly to incorporate $\Zbf$ and $\gammabf$. Also, we only consider GLM with a univariate response, whereas extensions to more complex models such as quasi-likelihood models and multivariate responses are straightforward.

For generality in terms of penalty, we consider the spectral regularization problem
\begin{eqnarray}
    \min_{\Bbf} \, h(\Bbf) = \ell(\Bbf) + J(\Bbf), \label{eqn:matrix-spectral-regularization}
\end{eqnarray}
where $\ell(\Bbf)$ is a loss function; for the GLM, we use the negative log-likelihood as the loss. $J(\Bbf) = f \circ \sigma(\Bbf)$, where $f: \real{q} \to \real{}$ is a function of the singular values of $\Bbf$. The choice $f(\wbf) = \lambda \sum_{j=1}^{q} |w_j|$ and the least squares loss corresponds to the special case of the nuclear norm regularization problem we considered in the Introduction. In general, for sparsity of the spectrum, $f$ takes the general form
\begin{eqnarray*}
  f(\wbf) = \sum_{j=1}^{q} P_\eta(|w_j|, \lambda),
\end{eqnarray*}
where $P$ is a scalar penalty function, $\eta$ is the parameter indexing the penalty family, and $\lambda$ is the tuning constant. We list some commonly used penalty functions below.
\begin{itemize}
\item Power family \citep{FrankFriedman93Bridge}
\begin{align*}
    P_\eta(|w|,\lambda) &= \lambda |w|^\eta, \hspace{.2in} \eta \in (0,2].
\end{align*}
Two important special cases of this family are the lasso penalty when $\eta=1$ \citep{Tibshirani96Lasso,ChenDonohoSaunders01BasisPursuit} and the ridge penalty when $\eta=2$ \citep{HoerlKennard70Ridge}.

\item Elastic net \citep{ZouHastie05Enet}
\begin{align*}
    P_\eta(|w|, \lambda) &= \lambda \left\{ (\eta-1) w^2/2 + (2-\eta) |w| \right\}, \hspace{.2in} \eta \in [1,2].
\end{align*}
Varying $\eta$ from 1 to 2 bridges the lasso to the ridge penalty.

\item Log penalty \citep{Candes08ReWtL1,Armagan11Pareto}
\begin{align*}
    P_\eta(|w|, \lambda) &= \lambda \ln (\eta + |w|), \hspace{.2in} \eta>0.
\end{align*}

\item SCAD \citep{FanLi01SCAD}, in which the penalty is defined via its partial derivative
\begin{align*}
    \frac{\partial}{\partial |w|} P_{\eta}(|w|, \lambda) &= \lambda \left\{ 1_{\{|w| \le \lambda\}} + \frac{(\eta \lambda - |w|)_+}{(\eta-1)\lambda} 1_{\{|w| > \lambda\}} \right\}, \hspace{.2in} \eta>2.
\end{align*}
Integration shows SCAD as a natural quadratic spline with knots at $\lambda$ and $\eta \lambda$
\begin{align*}
    P_{\eta}(|w|, \lambda) &= \begin{cases}
        \lambda|w| & |w|<\lambda  \\
        \lambda^2 + \frac{\eta \lambda(|w|-\lambda)}{\eta-1} - \frac{w^2-\lambda^2}{2(\eta-1)}   & |w| \in [\lambda, \eta\lambda]   \\
        \lambda^2 (\eta+1)/2  & |w|>\eta\lambda
    \end{cases}.
\end{align*}
For small signals $|w|<\lambda$, it acts as lasso; for large signals $|w|>\eta \lambda$, the penalty flattens and leads to the unbiasedness of the regularized estimate.

\item MC+ penalty \citep{Zhang10MCP}, which is similar to SCAD and is defined by the partial derivative
\begin{align*}
    \frac{\partial}{\partial |w|} P_{\eta}(|w|, \lambda) &= \lambda \left(1 - \frac{|w|}{\lambda \eta} \right)_+.
\end{align*}
Integration shows that the penalty function
\begin{align*}
    P_{\eta}(|w|, \lambda) &= \left(\lambda|w|-\frac{w^2}{2\eta} \right) 1_{\{|w|<\lambda \eta\}} + \frac{\lambda^2 \eta}{2} 1_{\{|w| \ge \lambda \eta\}}, \hspace{.2in} \eta>0,
\end{align*}
is quadratic on $[0,\lambda\eta]$ and flattens beyond $\lambda \eta$.
Varying $\eta$ from 0 to $\infty$ bridges hard thresholding ($\ell_0$ regression) to lasso ($\ell_1$) shrinkage.
\end{itemize}
We also comment that, besides the above sparsity penalties, other forms of regularization can be useful, depending on the scientific question of interest. For instance, the choice $f(\wbf) = \lambda \sum_{j=1}^{q-1} |w_j - w_{j+1}| = \lambda (w_1 - w_r)$ produces the regularization for the ``spiked" matrix model, i.e., matrices with clustered eigen-/singular values \citep{Johnstone01TW}.

Convexity is essential for studying convergence properties of optimization problems. We first state the necessary and sufficient condition for the convexity of the regularizer $J$. Its proof follows from the theory of spectral function \citep{BorweinLewis06ConvexAnalysisBook} and is given in the Appendix.
\begin{lemma}
\label{lem:J-subdiff}
The functional $J(\Bbf)=f \circ \sigma(\Bbf)$ is convex and lower semicontinuous if and only if $f$ is convex and lower semicontinuous. Furthermore, for a convex $f$, the subdifferential of $J$ at $\Bbf$, which admits singular value decomposition $\Ubf \mathrm{diag}(\bbf) \Vbf \trans$, is
\begin{eqnarray*}
    \partial J(\Bbf) = \partial (f \circ \sigma)(\Bbf) = \Ubf \mathrm{diag}[\partial f(\bbf)] \Vbf \trans.
\end{eqnarray*}
\end{lemma}
\noindent
Lemma \ref{lem:J-subdiff} immediately leads to the optimality condition when both loss and regularizer are convex.
\begin{theorem}
\label{thm:optimality}
When both the loss $\ell$ and $f$ are convex, all local minima of the regularized program \eqref{eqn:matrix-spectral-regularization} are global minimum and is unique if $\ell$ is strictly convex. A matrix $\Bbf = \Ubf \mathrm{diag}(\bbf) \Vbf \trans$ is a global minimum if and only if
\begin{eqnarray*}
    \zerobf_{p_1 \times p_2} \in \nabla \ell(\Bbf) + \Ubf \mathrm{diag}[\partial f(\bbf)] \Vbf \trans.
\end{eqnarray*}
\end{theorem}
\noindent
When either the loss $\ell$ or $f$ is non-convex, the regularized objective function \eqref{eqn:matrix-spectral-regularization} may be non-convex and there lacks an easy-to-check optimality condition.

\section{Estimation Algorithm}
\label{sec:algorithm}

We utilize the powerful Nesterov optimal gradient method \citep{Nesterov83NesterovI,Nesterov04ConvexOptmBook} for minimizing the non-smooth and possibly non-convex objective function \eqref{eqn:matrix-spectral-regularization}. We first state a matrix thresholding formula for spectral regularization, which forms the building blocks of the Nesterov algorithm.
\begin{proposition}
\label{prop:matrix-thresholding}
For a given matrix $\Abf$ with singular value decomposition $\Abf = \Ubf \mathrm{diag} (\abf) \Vbf \trans$, the optimal solution to
\begin{eqnarray*}
    \min_{\Bbf} \frac 12 \|\Bbf - \Abf\|_{\mathrm{F}}^2 + f \circ \sigma(\Bbf)
\end{eqnarray*}
shares the same singular vectors as $\Abf$ and its ordered singular values are the solution to
\begin{eqnarray*}
    \min_{\bbf} \frac 12 \|\bbf - \abf\|_2^2 + f(\bbf).
\end{eqnarray*}
\end{proposition}
\noindent
An immediate consequence of Proposition \ref{prop:matrix-thresholding} is the following well-known singular value thresholding formula for nuclear norm regularization \citep{CaiCandesShen10SVT}.
\begin{corollary}
\label{prop:matrix-thresholding}
For a given matrix $\Abf$ with singular value decomposition $\Abf = \Ubf \mathrm{diag} (\abf) \Vbf \trans$, the optimal solution to
\begin{eqnarray*}
    \min_{\Bbf} \frac 12 \|\Bbf - \Abf\|_{\mathrm{F}}^2 + \lambda \|\Bbf\|_*
\end{eqnarray*}
shares the same singular vectors as $\Abf$ and its singular values are $b_i = (a_i - \lambda)_+$.
\end{corollary}

Given the above matrix thresholding formula, we are ready to present the Nesterov algorithm for minimizing \eqref{eqn:matrix-spectral-regularization}. The Nesterov method has attracted increasing attention in recent years due to its efficiency in solving regularization problems \citep{BeckTebloulle09FISTA}. It resembles the classical gradient descent algorithm in that only the first order gradients of the objective function are utilized to produce next algorithmic iterate from current search point, and as such is simple to implement. It differs from the gradient descent algorithm by extrapolating the previous two algorithmic iterates to generate the next search point. This extrapolation step incurs trivial computational cost but improves the convergence rate dramatically. It has been shown to be optimal among a wide class of convex smooth optimization problems \citep{Nemirovski94,Nesterov04ConvexOptmBook}.

We summarize our Nesterov method for solving spectral regularization problem \eqref{eqn:matrix-spectral-regularization} in Algorithm \ref{algo:matrix-Nesterov}. %It can be considered as the analog to the FISTA algorithm for solving lasso penalized regression problem \citep{BeckTebloulle09FISTA}.
Each iteration consists of three steps: (i) predict a search point $\Sbf$ by a linear extrapolation from previous two iterates (line \ref{search-step} of Algorithm \ref{algo:matrix-Nesterov}), (ii) perform gradient descent from the search point $\Sbf$ possibly with Armijo type line search (lines \ref{eqn:graddes-start}-\ref{eqn:graddes-end}), and (iii) force the descent property of the next iterate (lines \ref{eqn:descheck-start}-\ref{eqn:descheck-end}).

\begin{algorithm}[t]
Initialize $\Bbf^{(0)}=\Bbf^{(1)}$, $\delta>0$, $\alpha^{(0)}=0$, $\alpha^{(1)}=1$ \;
\Repeat{objective value converges}{
$\Sbf^{(t)} \gets \Bbf^{(t)} + \left( \frac{\alpha^{(t-1)}-1}{\alpha^{(t)}} \right) (\Bbf^{(t)}-\Bbf^{(t-1)})$ \label{search-step} \;
\Repeat{$h(\Bbf_{\mathrm{temp}}) \le g(\Bbf_{\mathrm{temp}} \mid \Sbf^{(t)}, \delta)$ \label{eqn:graddes-end}}{ \label{eqn:graddes-start}
$\Abf_{\mathrm{temp}} \gets \Bbf^{(t)}- \delta \nabla \ell(\Sbf^{(t)})$ \;
Compute SVD $\Abf_{\mathrm{temp}} = \Ubf \text{diag}(\abf) \Vbf \trans$ \;
$\bbf \gets \text{argmin}_{\xbf} (2\delta)^{-1} \|\xbf - \abf\|_2^2 + f(\xbf)$ \;
$\Bbf_{\mathrm{temp}} \gets \Ubf \text{diag}(\bbf) \Vbf \trans$ \;
$\delta \gets \delta/2$ \;
}
\eIf{$h(\Bbf_{\mathrm{temp}}) \le h(\Bbf^{(t)})$ \label{eqn:descheck-start}}{
$\Bbf^{(t+1)} \gets \Bbf_{\mathrm{temp}}$ \;
}{
$\Bbf^{(t+1)} \gets \Bbf^{(t)}$ \;
} \label{eqn:descheck-end}
$\alpha^{(t+1)} \gets (1+\sqrt{1+(2\alpha^{(t)})^2})/2$ \label{eqn:update-alpha}
}
\caption{Nesterov method for spectral regularized matrix regression \eqref{eqn:matrix-spectral-regularization}.}
\vspace{0.1in}
\label{algo:matrix-Nesterov}
\end{algorithm}

In step (i), $\alpha^{(t)}$ is a scalar sequence that plays a critical role in the extrapolation. We update this sequence as in the original Nesterov method (line \ref{eqn:update-alpha} of Algorithm \ref{algo:matrix-Nesterov}), whereas other sequences, for instance $\alpha^{(t)}=(t-1)/(t+2)$, can also be used. In step (ii), the gradient descent is based on the first order approximation to the loss function at the current search point $\Sbf^{(t)}$,
\begin{eqnarray*}
    g(\Bbf|\Sbf^{(t)},\delta) &=& \ell(\Sbf^{(t)}) + \langle \nabla \ell(\Sbf^{(t)}), \Bbf - \Sbf^{(t)} \rangle + \frac{1}{2\delta} \|\Bbf - \Sbf^{(t)}\|_{\text{F}}^2 + J(\Bbf)    \\
    &=& \frac{1}{2\delta} \|\Bbf - [\Sbf^{(t)}-\delta \nabla \ell(\Sbf^{(t)})]\|_{\text{F}}^2 + J(\Bbf) + c^{(t)},
\end{eqnarray*}
where the constant $\delta$ is determined during the line search and the constant $c^{(t)}$ collects terms irrelevant to the optimization. The ``ridge" term $(2\delta)^{-1} \|\Bbf - \Sbf^{(t)}\|_{\text{F}}^2$ acts as a trust region and shrinks the next iterate towards $\Sbf^{(t)}$. If the loss function $\ell \in \mathcal{C}_{1,1}$, which denotes the class of functions that are convex, continuously differentiable and the gradient satisfies $\|\nabla \ell(\ubf) - \nabla \ell(\vbf)\| \le \mathcal{L}(\ell) \|\ubf - \vbf\|$ with a known gradient Lipschitz constant $\mathcal{L}(\ell)$ for all $\ubf, \vbf$, then $\delta$ is fixed at $\mathcal{L}(\ell)^{-1}$. In practice, the gradient Lipschitz constant is often unknown. Then $\delta$ is updated dynamically to capture the unknown $\mathcal{L}(\ell)$ using the classical Armijo line search rule \citep{NocedalWright06Book,Lange04Optm}. Solution to the surrogate function $g(\Bbf|\Sbf^{(t)},\delta)$ is obtained by Proposition \ref{prop:matrix-thresholding}. Singular value decomposition is performed on the intermediate matrix $\Abf_{\mathrm{temp}} = \Bbf^{(t)}-\delta \nabla \ell (\Sbf^{(t)})$. The next iterate $\Bbf^{(t+1)}$ shares the same singular vectors as $\Abf$ and its singular values $\bbf^{(t+1)}$ are determined by minimizing $(2\delta)^{-1} \|\bbf - \abf\|_2^2 + f(\bbf)$, where $\abf = \sigma(\Abf_{\mathrm{temp}})$. For a nuclear norm regularization $f(\wbf) = \lambda \sum_j |w_i|$, the solution is given by soft thresholding the singular values $b_i^{(t+1)} = (a_i-\lambda\delta)_+$. In this special case, only the top singular values/vectors need to be retrieved. The Lanczos method \citep{Golub96Book} is extremely efficient for this purpose. For a linear regularization function $f(\wbf) = \lambda \|\Dbf \wbf \|_1$ where $\Dbf$ has full column rank, reparameterization $\cbf = \Dbf \bbf$ turns the problem to:
$\min_{\cbf} \frac{1}{2\delta} \|(\Dbf\trans \Dbf)^{-1} \Dbf\trans \cbf - \abf\|_2^2 + \lambda \|\cbf\|_1,$
which is a standard lasso problem with many efficient solvers available. When $\Dbf$ does not have a full column rank, we append extra rows such that the expanded matrix, denoted by $\tilde \Dbf$, has full column rank and then solve the above lasso problem with $\tilde \Dbf$ and only part of $\cbf$ penalized.

For minimization of a smooth convex function $\ell$ in $\mathcal{C}_{1,1}$, it is well-known that the Nesterov method is optimal with the convergence rate at order $O(t^{-2})$, where $t$ indicates the iterate number. In contrast, the gradient descent has a slower convergence rate of $O(t^{-1})$. Our nuclear norm regularization problem \eqref{eqn:matrix-spectral-regularization} is non-smooth, but the same convergence result can be established, which is summarized in Theorem \ref{thm:convergencerate}. Its proof is omitted for brevity, while readers are referred to \citet{BeckTebloulle09FISTA}.
\begin{theorem}
\label{thm:convergencerate}
Suppose $\ell$ is continuously differentiable with a gradient Lipschitz constant $\mathcal{L}(\ell)$. Let $\Bbf^{(t)}$ be the iterates generated by the Nesterov method described in Algorithm \ref{algo:matrix-Nesterov}. Then the objective value $h(\Bbf^{(t)})$ monotonically converges. Furthermore, if $J$ is convex, then
\begin{eqnarray}
    h(\Bbf^{(t)})-h(\Bbf^*) \le \frac{4 \mathcal{L}(\ell) \|\Bbf^{(0)} - \Bbf^*\|_{\mathrm{F}}^2}{(t+1)^2}   \label{eqn:Nesterov-rate}
\end{eqnarray}
for all $t \ge 0$ and any minimum point $\Bbf^*$.
\end{theorem}

We make a few important remarks here. The first remark regards the monotonicity of the objective function during iterations. Because of the extrapolation step, the objective values of algorithmic iterates $f(\Bbf^{(t)})$ are not guaranteed to be monotonically decreasing. When the loss $\ell \in {\cal C}_{1,1}$ and the regularizer $J$ is convex, convergence of the objective values is guaranteed with the explicit convergence rate \eqref{eqn:Nesterov-rate}. Because of potential use of a non-convex $J$, we enforce monotonicity of algorithmic iterates (lines \ref{eqn:descheck-start}-\ref{eqn:descheck-end} in Algorithm \ref{algo:matrix-Nesterov}), which is essential for the convergence of at least the objective values. After each gradient descent step, if the new iterate fails to decrease the objective value, then the current iterate is same as the previous one. In other words, the next gradient descent is initiated from the previous iterate. Fortunately, the fast convergence rate \eqref{eqn:Nesterov-rate} still holds under the assumptions $\ell \in {\cal C}_{1,1}$ and $J$ is convex. See \citet{BeckTeboulle09FastGradient} for the argument.

The second remark is about the non-convex loss function. The Nesterov method and its convergence properties hinge upon convexity of the loss $\ell$. It covers many commonly used statistical models, including linear model and GLMs with canonical links. For GLM with non-canonical link, the loss function may be non-convex, which could cause trouble in the Nesterov method. The iteratively reweighted least squares strategy can be applied in this scenario. At each IWLS step, the Nesterov method is used to solve the penalized weighted least squares problem, which is convex.

The final remark is about an efficient way for estimating the Lipschitz constant $L$ for the GLM loss. Each step halving in the line search part of Algorithm \ref{algo:matrix-Nesterov} involves an expensive singular value decomposition. Therefore even a rough initial estimate of $L$ potentially cuts the computational cost significantly. Recall that a twice differentiable function $f$ is $L$-Lipschitz continuously differentiable if and only if $\vbf \trans d^2 f(\ubf) \vbf \le L \|\vbf\|_2^2$ for all $\vbf$. The Fisher information matrix of a GLM model with systematic part \eqref{eqn:tensor-reg-eta} is:
$\Ibf(\Bbf) = \mathrm{E} [d^2 \ell(\Bbf)] = \sum_{i=1}^n \omega_i (\mathrm{vec} \xbf_i)(\mathrm{vec} \xbf_i) \trans$,
where $\omega_i = \{\mu_i'(\eta_i) / \sigma_i\}^2$, $\eta_i$ is the systematic part, $\mu_i$ is the mean, and $\sigma_i^2$ is the variance corresponding to the $i$th observation. Then in light of the Cauchy-Schwartz inequality
\begin{eqnarray*}
    \vbf \trans \Ibf(\Bbf) \vbf = \sum_i \omega_i (\vbf \trans \vect \xbf_i)(\vbf \trans \vect \xbf_i) \trans \le \sum_i \omega_i \|\mathrm{vec} \xbf_i\|_2^2 \|\vbf\|_2^2 = \|\vbf\|_2^2 \left( \sum_i \omega_i \|\xbf_i\|_{\mathrm{F}}^2 \right),
\end{eqnarray*}
and thus an initial estimate of $L$ is given by $L \approx \sum_i \{\mu_i'(\eta_i)\}^2 / \sigma_i^2 \; \|\Xbf_i\|_{\mathrm{F}}^2$.

\section{Degrees of Freedom}
\label{sec:dof}

In this section, we address the problem of choosing the tuning parameter $\lambda$ that yields the best model along the regularization path according to certain criteria. Cross validation is commonly used for parameter tuning in practice. However, for large data, it may incur considerable computation burden. There exist computationally attractive alternatives, such as Akaike information criterion (AIC) \citep{Akaike74AIC} and Bayesian information criterion (BIC) \citep{Schwartz78BIC}, which often yield performance comparable to cross validation in practice.

Consider a normal model under the GLM \eqref{eqn:tensor-reg-eta}. For simplicity, we again drop the covariate vector $\Zbf$:
\begin{eqnarray}
    Y = \langle \Xbf, \Bbf \rangle + \epsilon \label{eqn:normal-model}
\end{eqnarray}
where $\epsilon$ is a normal error with mean zero and variance $\sigma^2$. Let $y_i$ denote the $i$th observation of $Y$, and $\hat{y}_i(\lambda)$ denote the estimated response under a given tuning parameter $\lambda$ from the minimization of \eqref{eqn:matrix-spectral-regularization}. Then for this normal model, AIC and BIC are defined by
\begin{eqnarray*}
    \mathrm{AIC}(\lambda) &=& \frac{\sum_{i} \{y_i - \hat{y}_i(\lambda)\}^2}{\sigma^2} + 2 \mathrm{df}(\lambda)    \\
    \mathrm{BIC}(\lambda) &=& \frac{\sum_{i} \{y_i - \hat{y}_i(\lambda)\}^2}{\sigma^2} + \ln (n) \mathrm{df}(\lambda).
\end{eqnarray*}
In applications, the variance $\sigma^2$ is often unknown but can be estimated from the fitted value by least squares estimation. An essential element in the above model selection criteria is the effective degrees of freedom $\mathrm{df}(\lambda)$ of the selected model. Using Stein's theory of unbiased risk estimation \citep{Stein81MVN}, \citet{Efron04CV} showed that
\begin{eqnarray*}
    \mathrm{df}(\lambda) = \mathrm{E} \left\{ \mathrm{tr} \left( \frac{\partial \hat\ybf}{\partial \ybf} \right) \right\} = \frac{1}{\sigma^2} \sum_{i=1}^n \mathrm{cov}(\hat y_i(\lambda),y_i)
\end{eqnarray*}
with expectation taken with respect to $Y$, $\ybf = (y_1,\ldots,y_n)\trans$, and $\hat\ybf = (\hat y_1(\lambda),\ldots, \hat y_n(\lambda))\trans$. This formulation has been productively used to derive the degrees of freedom estimate in least angle regression \citep{EfronHastieIainTibshirani04LARS},  lasso \citep{ZouHastieTibshirani07LassoDF}, group penalized regression \citep{YuanLin06GroupLasso}, and sign-coherent group penalized regression \citep{Chiquet11CoopGpLasso}.

We derive a degrees of freedom estimate for the nuclear norm regularized estimate under normal model. For an orthonormal design, i.e., $\Xibf \trans \Xibf = \Ibf_{p_1p_2}$ with the matrix $\Xibf$ having rows $(\mathrm{vec} \xbf_i) \trans$, the derived estimate is unbiased for the true degrees of freedom. In practice, it yields results comparable to cross validation even for non-orthogonal designs. The technical proof is relegated to the Appendix.
\begin{theorem}
\label{thm:dof}
Assume that the data is generated from model \eqref{eqn:normal-model} with $\mathrm{vec} \xbf_i$ orthonormal. Consider the nuclear norm regularized estimate
\begin{eqnarray*}
\widehat \Bbf_\lambda = \mathrm{argmin}_{\Bbf} \frac 12 \sum_i (y_i - \langle \Xbf_i, \Bbf \rangle)^2 + \lambda \|\Bbf\|_*
\end{eqnarray*}
with singular values $\sigma(\widehat \Bbf_\lambda) = (b_1(\lambda), \ldots, b_q(\lambda))$ where $q = \min \{p_1,p_2\}$. Let $\widehat \Bbf_{\mathrm{LS}}$ be the usual least squares estimate and assume that it has distinct positive singular values $\sigma_1 > \cdots > \sigma_q >0$. With the convention $\sigma_i =0$ for $i>q$, the following expression is an unbiased estimate of the degree of freedom of the regularized fit
\begin{eqnarray*}
    \widehat{\mathrm{df}}(\lambda) = \sum_{i=1}^q 1_{\{b_i(\lambda)>0\}} \left( 1 + \sum_{1 \le j \le p_1, j \ne i} \frac{\sigma_i(\sigma_i-\lambda)}{\sigma_i^2 - \sigma_j^2} + \sum_{1 \le j \le p_2, j \ne i} \frac{\sigma_i(\sigma_i-\lambda)}{\sigma_i^2 - \sigma_j^2} \right).
\end{eqnarray*}
\end{theorem}
This formula for the degrees of freedom is interesting in several aspects. First it does not involve any information on the singular vectors of the least squares estimate,  but only requires singular values. Second, $\widehat{\mathrm{df}}(\lambda)$ is continuous in $\lambda$, in contrast to the piecewise constant degrees of freedom estimate for the classical lasso \citep{ZouHastieTibshirani07LassoDF}. Third, at $\lambda=0$, $\widehat \Bbf_{0} = \widehat \Bbf_{\mathrm{LS}}$ almost surely has a full rank and the degrees of freedom is
\begin{eqnarray*}
    & & \sum_{i=1}^q \left( 1 + \sum_{1\le j \le p_1, j \ne i} \frac{\sigma_i^2}{\sigma_i^2 - \sigma_j^2} + \sum_{1\le j \le p_2, j \ne i} \frac{\sigma_i^2}{\sigma_i^2 - \sigma_j^2} \right) \\
    &=& q + q(q-1) + q(p_1+p_2-2q)  = p_1p_2,
\end{eqnarray*}
which is exactly the number of parameters without any regularization.

\begin{figure}
\centering
\includegraphics[width=3.5in]{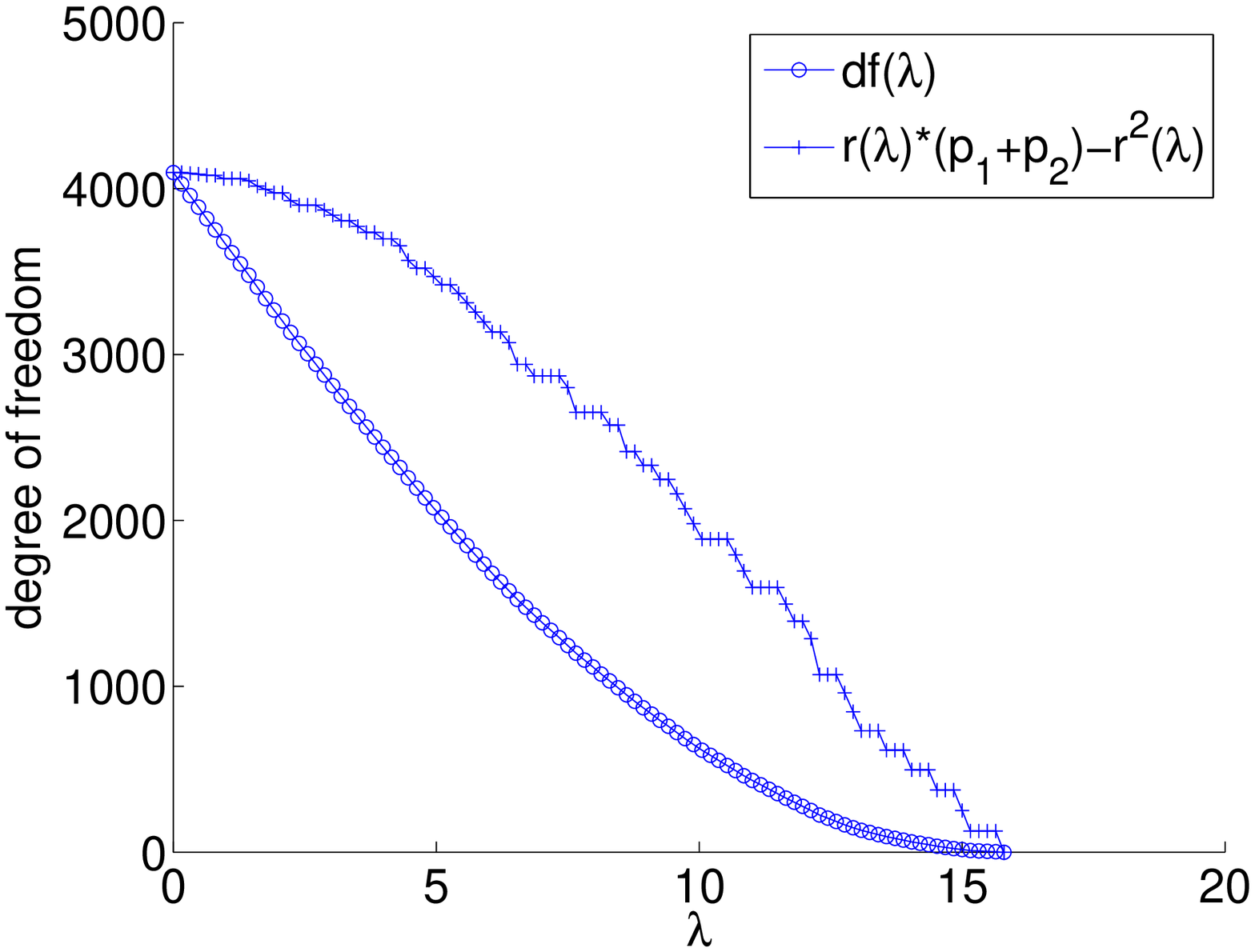}
\caption{Degrees of freedom estimate $\widehat{\mathrm{df}}(\lambda)$ versus the number of parameters in the estimated model $\widehat \Bbf_\lambda$ with a $\widehat \Bbf_{\mathrm{LS}} \in \real{64 \times 64}$.}
\label{fig:dof}
\end{figure}

Figure \ref{fig:dof} plots the estimated degrees of freedom $\widehat{\mathrm{df}}(\lambda)$ for a $\widehat \Bbf_{\mathrm{LS}} \in \real{64 \times 64}$, as well as the na\"{i}ve count of the number of free parameters in the fitted matrix parameter $\widehat \Bbf_\lambda$ of rank $r(\lambda)$, which equals $r(\lambda) (p_1+p_2) - r^2(\lambda)$. The estimated degrees of freedom appears smaller than the na\"{i}ve count, reflecting the overwhelming shrinkage effect over model searching. At $\lambda=0$, it coincides with the number of matrix elements $64^2 = 4096$, reflecting the effect of no shrinkage.

Finally we notice that the degrees of freedom estimate in Theorem \ref{thm:dof} is limiting as it requires existence of the least squares estimate $\widehat \Bbf_{\mathrm{LS}}$ which is not true when $n< p_1 p_2$. In this case we may use a ridge estimate $\widehat \Bbf_{\mathrm{ridge}(\tau)}$ where
\begin{eqnarray*}
    \mathrm{vec} \widehat \Bbf_{\mathrm{ridge}(\tau)} = (\Xibf \trans \Xibf + \tau \Ibf_{p_1p_2})^{-1} \Xibf \trans \ybf,
\end{eqnarray*}
which always exists and is unique. Assume that $\widehat \Bbf_{\mathrm{ridge}(\tau)}$ admits a singular value decomposition $\widehat \Bbf_{\mathrm{ridge}(\tau)} = \Ubf \mathrm{diag}(\sigmabf) \Vbf \trans$. The following degree of freedom formula
\begin{eqnarray*}
    \widehat{\mathrm{df}}(\tau) & = & \sum_{i=1}^q 1_{\{b_i(\tau)>0\}} \left[1 + \frac{1}{1+\tau} \sum_{1 \le j \le p_1, j \ne i} \frac{\sigma_i\{(1+\tau)\sigma_i-\lambda\}}{\sigma_i^2 - \sigma_j^2} \right. \\
    & & \qquad \qquad \qquad \quad \left. + \frac{1}{1+\tau} \sum_{1 \le j \le p_2, j \ne i} \frac{\sigma_i\{(1+\tau)\sigma_i-\lambda\}}{\sigma_i^2 - \sigma_j^2} \right]
\end{eqnarray*}
generalizes Theorem \ref{thm:dof} and is unbiased for the true degree of freedom under the same assumptions as Theorem \ref{thm:dof}. Its proof is given in the Appendix.

\section{Numerical Examples}
\label{sec:numerics}

We have conducted intensive numerical studies with two aims: first, we investigate the empirical performance of the proposed spectral regularized regression with matrix covariates, and second, we compare with the corresponding classical regularization solutions. Four methods are under comparison: a matrix regression with nuclear norm regularization (since it takes the form $f(\wbf) = \lambda \sum_{j=1}^{q} |w_j|$ in the regularization problem \eqref{eqn:matrix-spectral-regularization}, we call this solution matrix lasso, or simply, m.lasso), a usual vector regression after vectorizing the matrix covariate with a lasso penalty (lasso), a matrix regression with power spectral regularization (matrix power, or simply, m.power), and the corresponding vector regression with a power penalty (power). For the power penalty, we fixed the coefficient $\eta = 0.5$. Our goal here is not to best tune $\eta$. Instead, we examine this penalty since it yields a nearly unbiased estimate. We also examined another unbiased penalty, SCAD, which yields very similar results as power $\eta=0.5$, and thus its results are not reported here for brevity. We also note that the lasso penalty is a convex penalty, while the power $\eta=0.5$ is non-convex. We summarize our findings in three examples. First, we elaborated on the illustrative example by examining a number of different geometric and natural shapes. Second, we generated some synthetic data and compared different regularization solutions under varying ranks and sparsity. Lastly, we revisited the motivating electroencephalography (EEG) data analysis mentioned in the Introduction.

\begin{figure}
\centering
\begin{tabular}{c}
\includegraphics[width=3.8in]{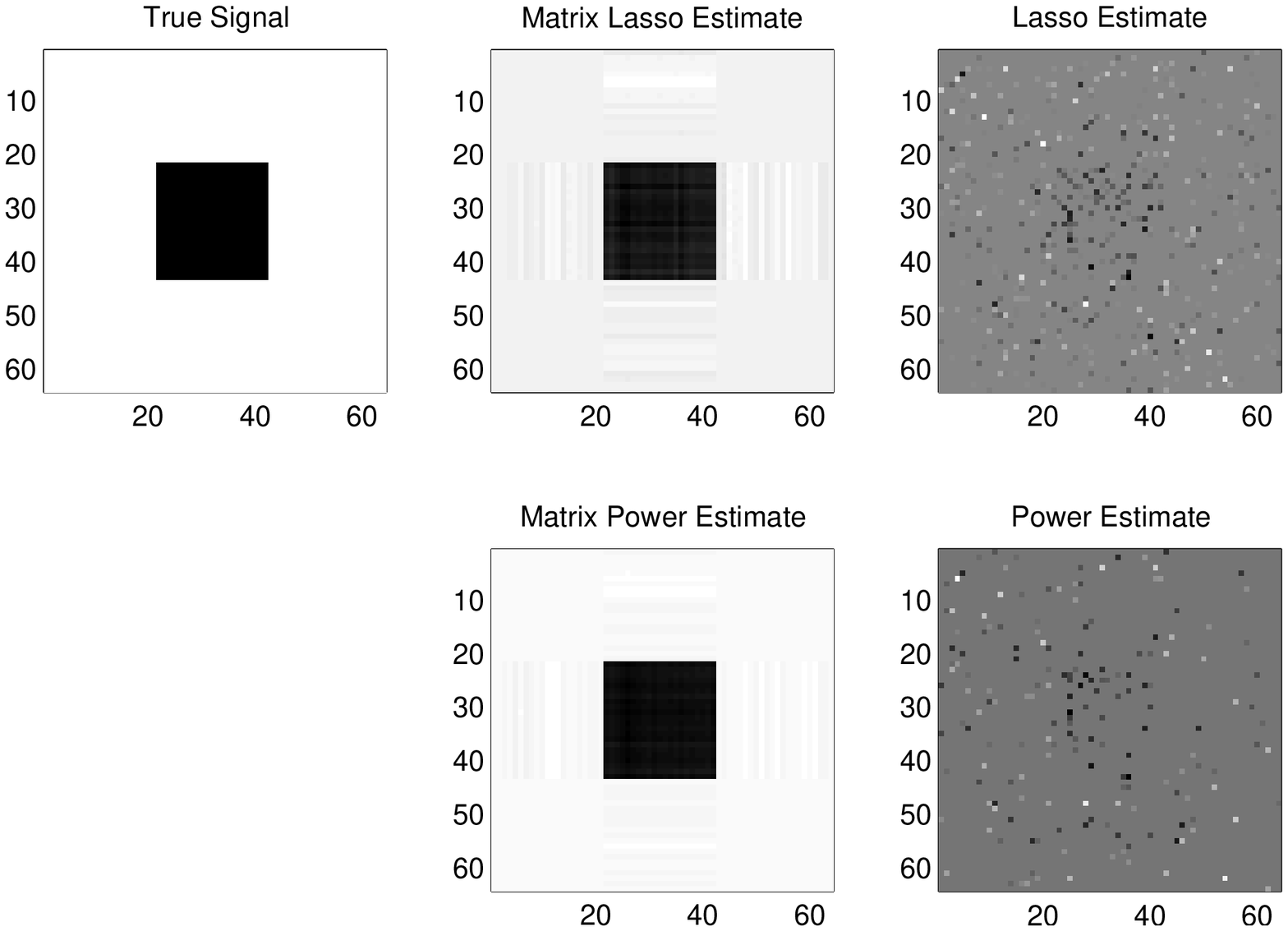} \\
\includegraphics[width=3.8in]{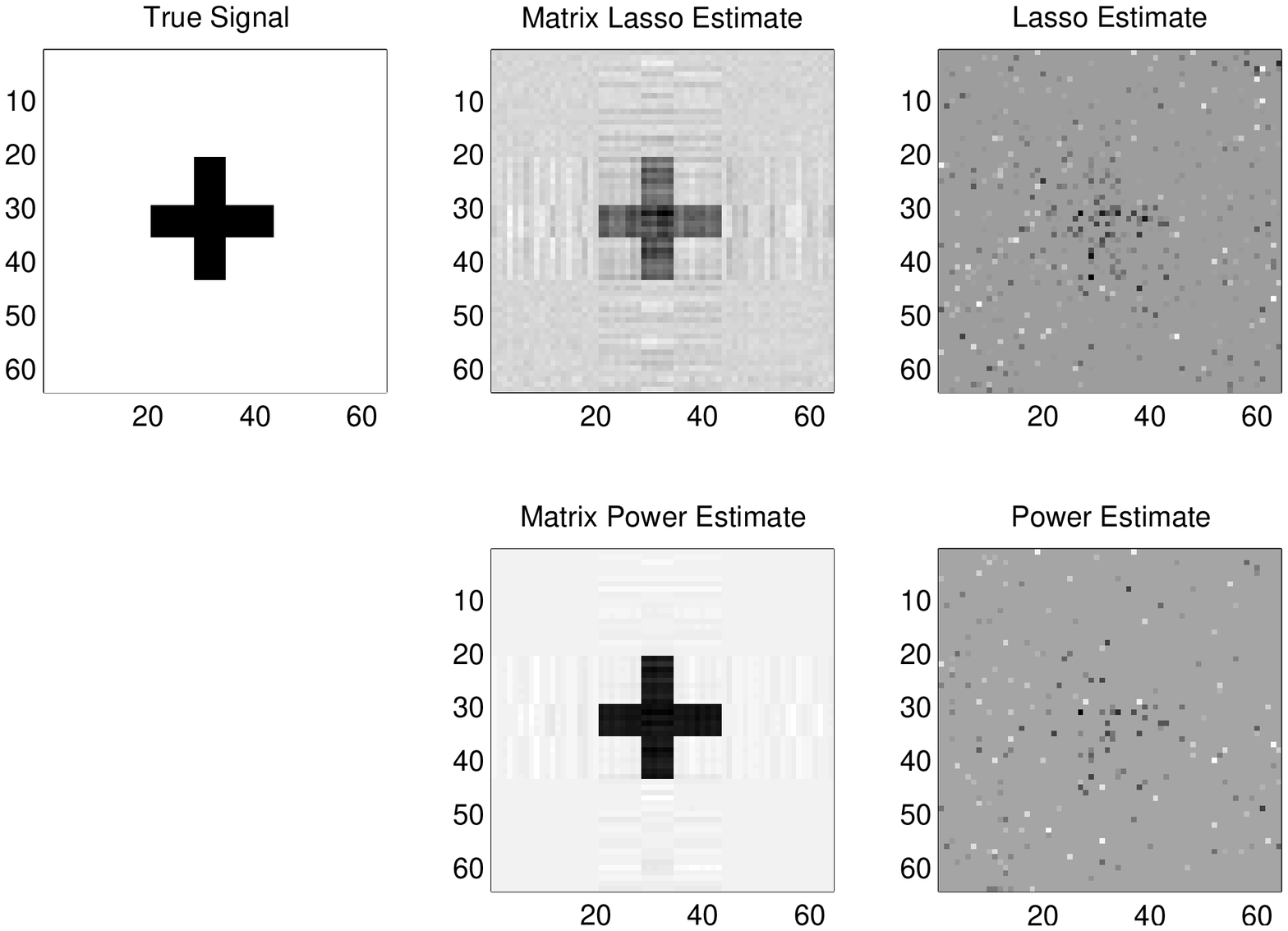} \\
\includegraphics[width=3.8in]{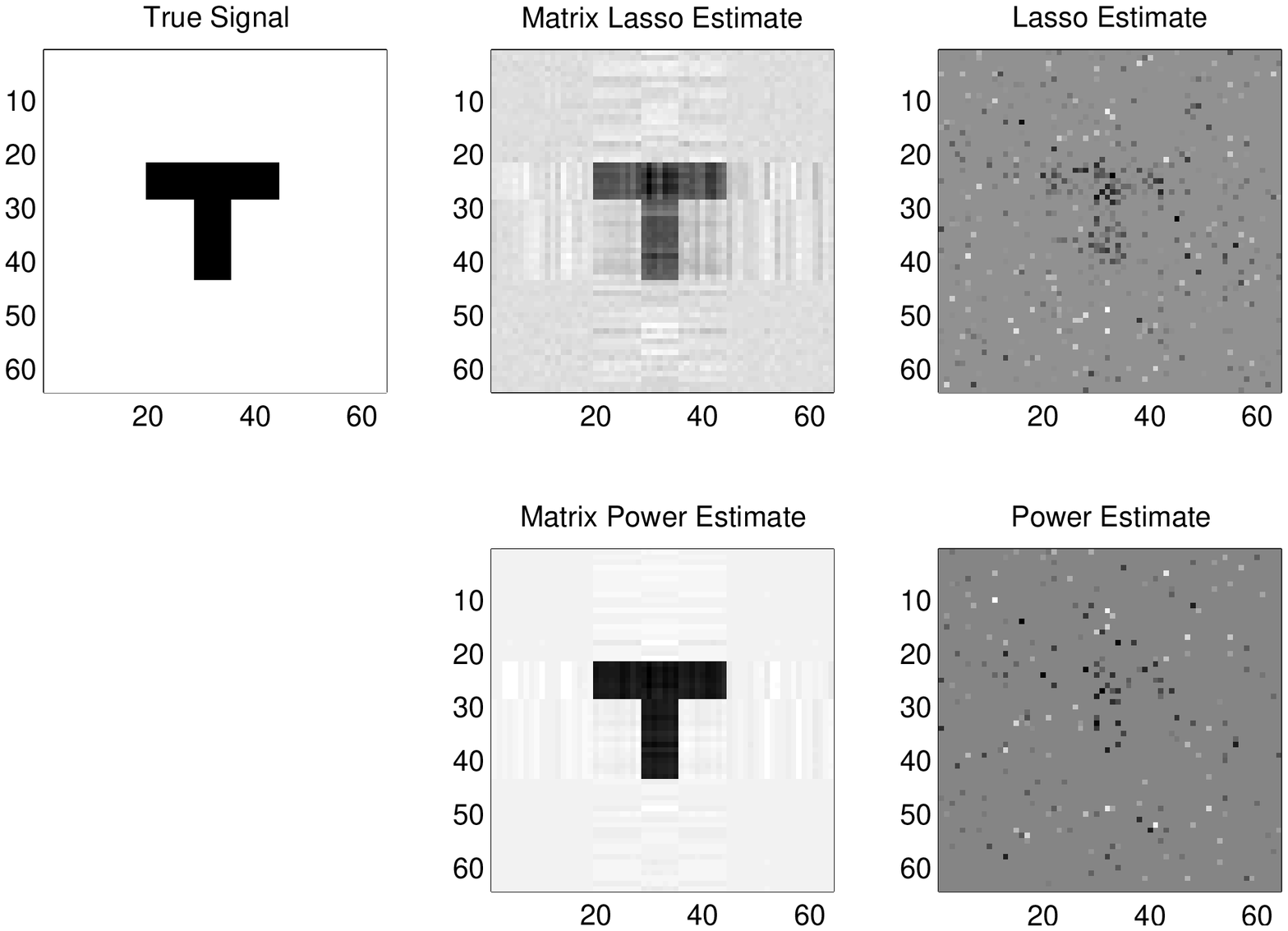} \\
\end{tabular}
\caption{Comparison of the matrix and vector version of regularized estimators for low rank signals.}
\label{fig:lowrank}
\end{figure}

\begin{figure}
\centering
\begin{tabular}{c}
\includegraphics[width=3.8in]{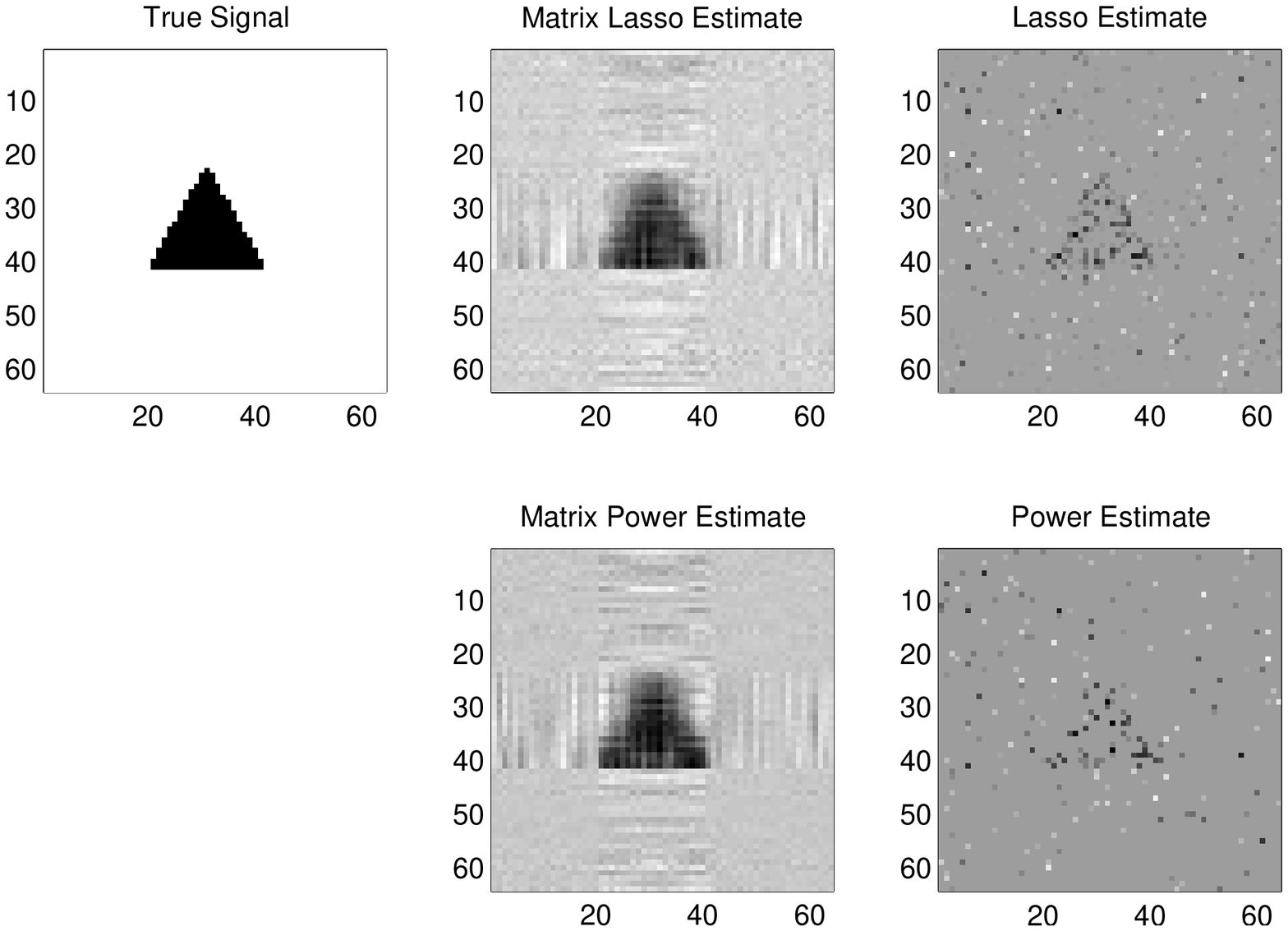} \\
\includegraphics[width=3.8in]{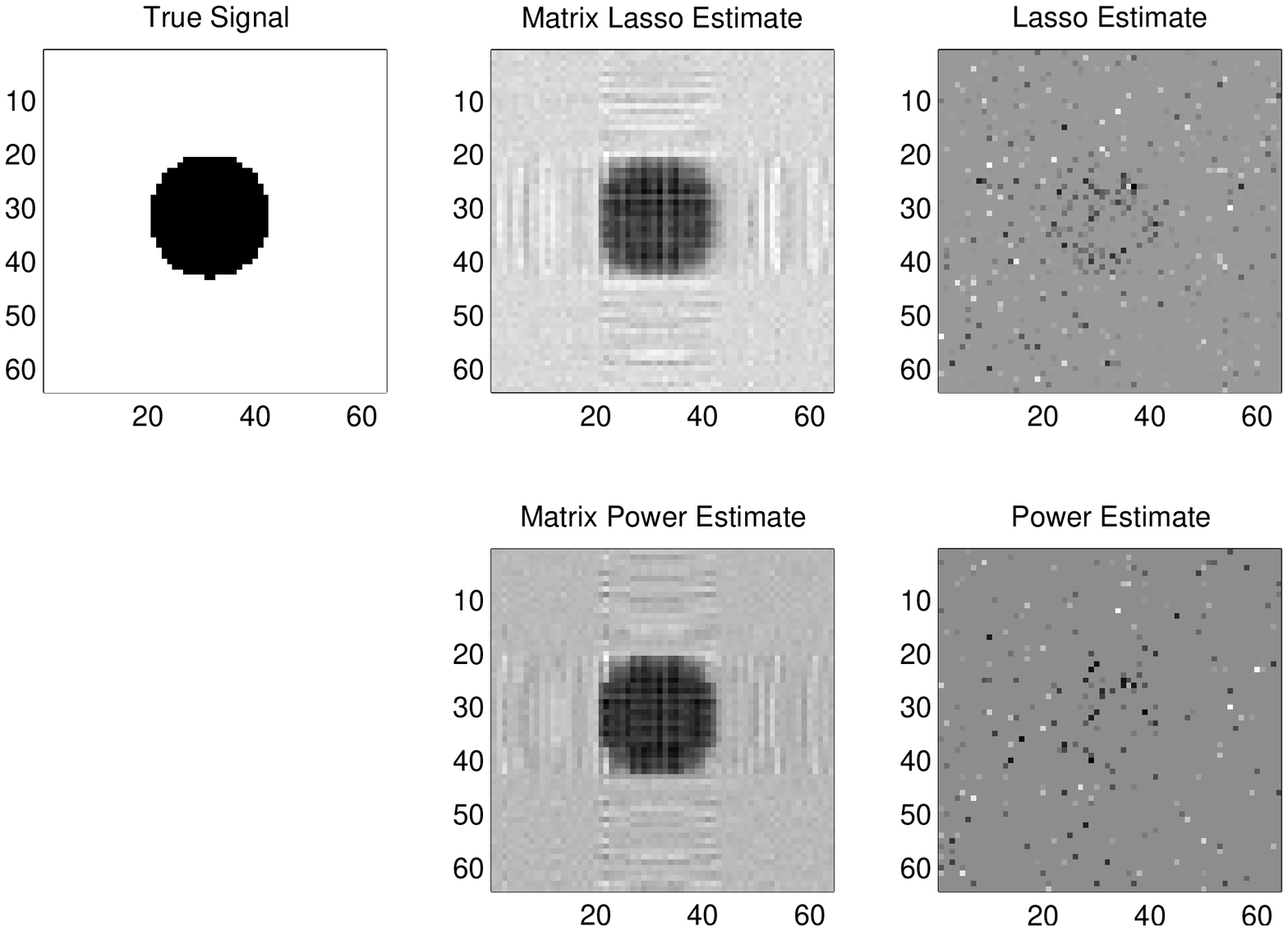} \\
\includegraphics[width=3.8in]{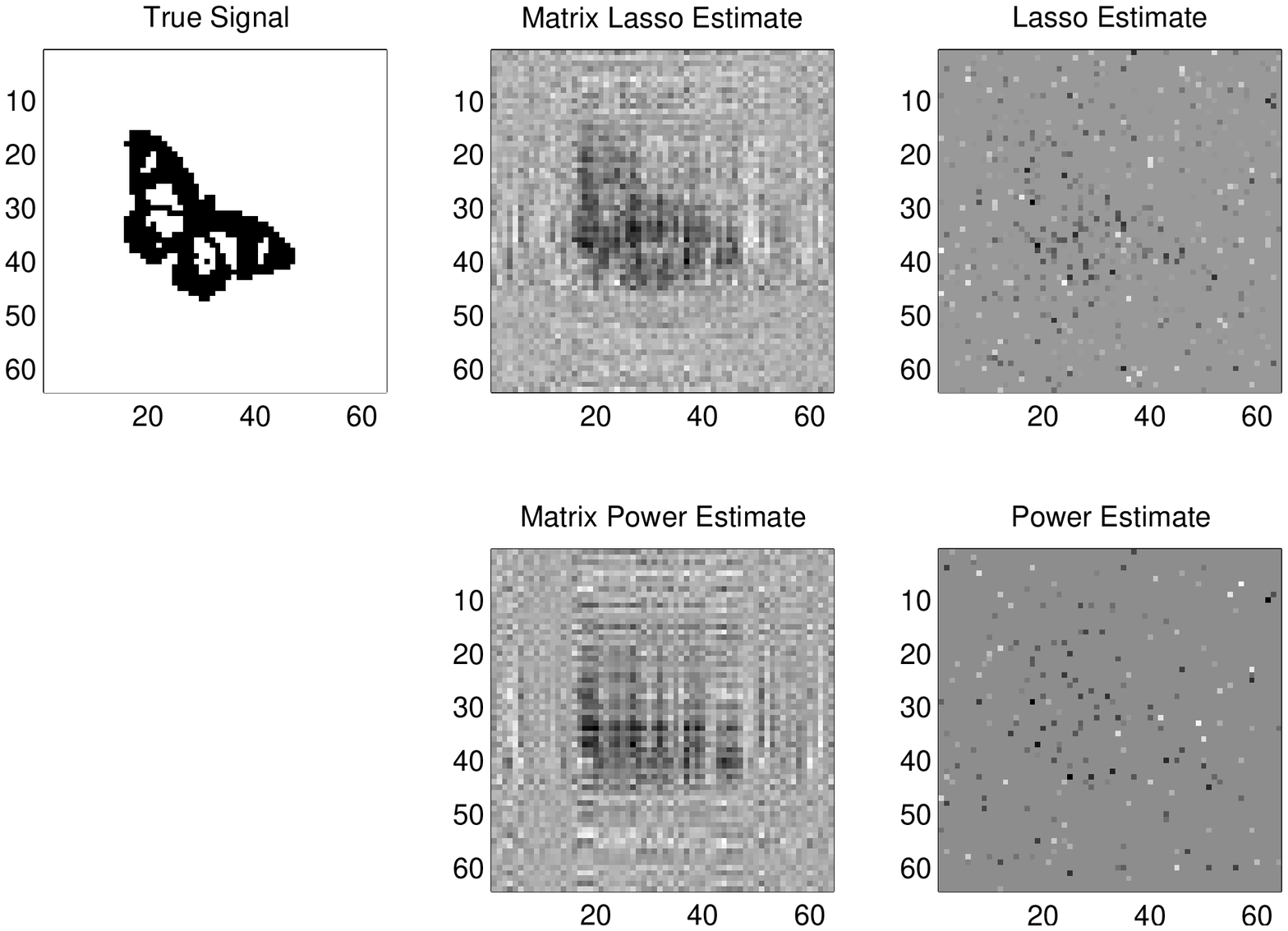} \\
\end{tabular}
\caption{Comparison of the matrix and vector version of regularized estimators for high rank signals.}
\label{fig:highrank}
\end{figure}

\medskip
\noindent
\textbf{Example 1: 2D Shapes.} We elaborated on the illustrative example in the Introduction, by employing the same model setup, but examining a variety of signal shapes. We presented the true signal followed by the estimates from the four aforementioned regularization methods, where the regularization parameter $\lambda$ was tuned by BIC. The signals of square, cross, T-shape are given in Figure \ref{fig:lowrank}, where the true signal is of a low rank structure (rank 1 or 2). The signals of triangle, disk and butterfly are given in Figure \ref{fig:highrank}, where the true signal is not of an exact low rank structure, however, can be well approximated by so \citep{ZhouLiZhu12CPTensor}. It is seen that the matrix version of regularized estimators clearly outperform their vector version counterparts, for both lasso and power penalties. Comparing matrix lasso with matrix power, the two yield comparable results, while the former is better for the high rank signals, and the latter is better for the low rank signals. We will further verify this observation in the next simulation example.

\begin{table}[t]
\caption{Parameter estimation of a normal model. Reported are the mean and standard deviation (in the parenthesis) of the RMSE for $\Bbf$ out of 100 data replications.}
\vspace{0.1in}
\centerline{
\begin{tabular}{clcccc}
\toprule
Sparsity & & \multicolumn{4}{c}{Rank}  \\ \cmidrule{3-6}
$s$  &  Method  & $R=1$ & $R=5$ & $R=10$ & $R=20$ \\ \midrule
$1\%$     & m.lasso   & 0.031 (0.006) & 0.074 (0.016) & 0.086 (0.015) & 0.090 (0.013) \\	
                 & lasso       & 0.022 (0.017) & 0.022 (0.010) & 0.023 (0.013) & 0.022 (0.006) \\	
                 & m.power & 0.010 (0.001) & 0.063 (0.021) & 0.097 (0.021) & 0.107 (0.017) \\	
                 & power     & {\bf 0.009} (0.025) & {\bf 0.005} (0.001) & {\bf 0.007} (0.017) & {\bf 0.006} (0.001) \\ \hline	
$5\%$     & m.lasso   & 0.044 (0.007) & 0.172 (0.021) & {\bf 0.199} (0.022) & {\bf 0.212} (0.020) \\	
                 & lasso       & 0.212 (0.049) & 0.214 (0.035) & 0.215 (0.032) & 0.217 (0.026) \\	
                 & m.power & {\bf 0.010} (0.001) & {\bf 0.171} (0.029) & 0.236 (0.029) & 0.254 (0.024) \\	
                 & power     & 0.270 (0.060) & 0.271 (0.042) & 0.272 (0.039) & 0.275 (0.032) \\ \hline	
$10\%$   & m.lasso   & 0.051 (0.008) & {\bf 0.250} (0.028) & {\bf 0.288} (0.023) & {\bf 0.309} (0.023) \\	
                 & lasso       & 0.320 (0.046) & 0.343 (0.040) & 0.343 (0.034) & 0.345 (0.030) \\	
                 & m.power & {\bf 0.009} (0.001) & 0.258 (0.041) & 0.351 (0.033) & 0.375 (0.032) \\	
                 & power     & 0.399 (0.056) & 0.427 (0.050) & 0.428 (0.044) & 0.426 (0.036) \\ \hline	
$20\%$   & m.lasso   & 0.062 (0.011) & {\bf 0.350} (0.031) & {\bf 0.418} (0.031) & {\bf 0.450} (0.031) \\	
                 & lasso       & 0.475 (0.050) & 0.523 (0.050) & 0.539 (0.049) & 0.534 (0.044) \\	
                 & m.power & {\bf 0.009} (0.001) & 0.354 (0.041) & 0.511 (0.039) & 0.562 (0.044) \\	
                 & power     & 0.585 (0.062) & 0.645 (0.060) & 0.667 (0.060) & 0.663 (0.055) \\ \hline	
$50\%$   & m.lasso   & 0.087 (0.015) & 0.556 (0.030) & {\bf 0.700} (0.044) & {\bf 0.792} (0.040) \\	
                 & lasso       & 0.775 (0.044) & 1.054 (0.081) & 1.112 (0.089) & 1.130 (0.069) \\	
                 & m.power & {\bf 0.010} (0.001) & {\bf 0.493} (0.042) & 0.760 (0.055) & 0.903 (0.062) \\	
                 & power     & 0.952 (0.057) & 1.300 (0.107) & 1.365 (0.113) & 1.388 (0.082) \\	
\bottomrule
\end{tabular}
\label{tab:norm-estB}
}
\end{table}

\begin{table}[t]
\caption{Prediction of a normal model. Reported are the mean and standard deviation (in the parenthesis) of the RMSE for $y$ out of 100 data replications.}
\vspace{0.1in}
\centerline{
\begin{tabular}{clcccc}
\toprule
Sparsity & & \multicolumn{4}{c}{Rank}  \\ \cmidrule{3-6}
$s$  &  Method  & $R=1$ & $R=5$ & $R=10$ & $R=20$ \\ \midrule
$1\%$     & m.lasso    & 1.707 (0.195) & 4.597 (1.091) & 5.525 (1.009) & 5.865 (0.826) \\	
                 & lasso        & 1.610 (0.978) & 1.544 (0.487) & 1.622 (0.762) & 1.530 (0.233) \\	
                 & m.power  & {\bf 1.175} (0.042) & 3.996 (1.164) & 5.534 (1.047) & 5.992 (0.874) \\	
                 & power      & 1.236 (1.240) & {\bf 1.133} (0.742) & {\bf 1.168} (0.962) & {\bf 1.066} (0.039) \\ \hline
$5\%$     & m.lasso    & 2.278 (0.309) & 11.031 (1.418) & {\bf 12.837} (1.578) & {\bf 13.754} (1.352) \\	
                 & lasso        & 13.488 (2.923) & 13.625 (2.115) & 13.688 (2.047) & 13.855 (1.609) \\	
                 & m.power  & {\bf 1.177} (0.043) & {\bf 10.253} (1.479) & 13.013 (1.593) & 14.118 (1.407) \\	
                 & power      & 14.713 (3.384) & 14.877 (2.109) & 14.933 (2.018) & 15.113 (1.597) \\ \hline	
$10\%$    & m.lasso    & 2.542 (0.420) & 16.101 (1.940) & {\bf 18.590} (1.568) & {\bf 19.810} (1.514) \\	
                 & lasso        & 19.601 (2.661) & 21.257 (2.488) & 21.165 (2.052) & 21.187 (1.742) \\	
                 & m.power  & {\bf 1.179} (0.045) & {\bf 15.386} (2.161) & 18.840 (1.588) & 20.239 (1.597) \\	
                 & power      & 21.138 (2.879) & 22.796 (2.782) & 22.708 (2.323) & 22.546 (1.776) \\ \hline	
$20\%$    & m.lasso    & 3.188 (0.655) & 22.723 (2.157) & {\bf 26.783} (2.264) & {\bf 28.874} (2.256) \\	
                 & lasso        & 28.687 (3.017) & 31.862 (3.219) & 32.588 (3.182) & 32.449 (2.767) \\	
                 & m.power  & {\bf 1.169} (0.042) & {\bf 21.401} (2.309) & 27.103 (2.171) & 29.271 (2.189) \\	
                 & power      & 30.879 (3.284) & 34.350 (3.495) & 35.013 (3.600) & 34.998 (2.994) \\ \hline	
$50\%$    & m.lasso    & 4.566 (1.026) & 35.651 (2.153) & 45.132 (2.989) & 50.916 (3.054) \\	
                 & lasso        & 45.815 (2.749) & 62.834 (4.896) & 66.490 (5.523) & 67.478 (4.444) \\	
                 & m.power  & {\bf 1.180} (0.047) & {\bf 29.638} (2.079) & {\bf 42.104} (2.381) & {\bf 48.892} (2.945) \\	
                 & power      & 49.798 (3.091) & 68.073 (5.719) & 71.626 (6.063) & 72.729 (4.986) \\	
\bottomrule
\end{tabular}
\label{tab:norm-pred}
}
\end{table}

\medskip
\noindent
\textbf{Example 2: Synthetic Data.} We consider a class of synthetic signals to compare various regularization methods under different ranks and sparsity levels. Specifically, we generated the matrix covariates $\Xbf$ of size $64 \times 64$ and the $5$-dimensional vector covariates $\Zbf$, both of which consist of independent standard normal entries. We set the sample size $n = 500$, whereas the number of parameters is $64 \times 64 + 5 = 4,101$. We set $\gammabf = (1, \ldots, 1)\trans$, and generated the true array signal as $\Bbf = \Bbf_1 \Bbf_2 \trans$, where $\Bbf_d \in \real{p_d \times R}$, $d=1,2$. $R$ controls the rank of the generated signal. Moreover, each entry of $\Bbf$ is $0/1$, and the percentage of non-zero entries is controlled by a sparsity level constant $s$. That is, each entry of $\Bbf_d$ is a Bernoulli with probability of one equal to $\sqrt{1 - (1 - s)^{1/R}}$. We varied the rank $R=1,5,10$ and $20$, and the level of (non)sparsity $s = 0.01, 0.05, 0.1,0.2$ and $0.5$. (So $s=0.05$ means about $5\%$ of entries of $\Bbf$ are ones and the rest are zeros.) We generated both a normal and a binomial response $Y$ with the systematic part as in \eqref{eqn:tensor-reg-eta}, $\mu = \eta$ for the normal model, and $\mu = 1/\{1 + \exp(-\eta)\}$ for the binomial.

We evaluated the performance of each method from two aspects: parameter estimation and prediction. For the former, we employed BIC for parameter tuning and the root mean squared error (RMSE) as the evaluation criterion. For the latter, we used an independent validation data set to tune the regularization parameter and a testing data set to evaluate the prediction error, which is measured by the RMSE of the response for the normal case, and the mis-classification error rate for the binomial case. Those are all common practices in the literature. We replicated the experiment 100 times, and report the mean and standard deviation of the corresponding criterion out of 100 replications in Tables \ref{tab:norm-estB} -- \ref{tab:bino-pred}. The best outcomes are highlighted in bold face. Since the RMSE for the vector-valued coefficient $\gammabf$ shows the same qualitative pattern as that for the array coefficient $\Bbf$, we only present the results for $\Bbf$ for brevity.

\begin{table}[t]
\caption{Parameter estimation of a binomial model. Reported are the mean and standard deviation (in the parenthesis) of the RMSE for $\Bbf$ out of 100 data replications.}
\vspace{0.1in}
\centerline{
\begin{tabular}{clcccc}
\toprule
Sparsity & & \multicolumn{4}{c}{Rank}  \\ \cmidrule{3-6}
$s$  &  Method  & $R=1$ & $R=5$ & $R=10$ & $R=20$ \\ \midrule
$1\%$     & m.lasso    & 0.087 (0.023) & 0.094 (0.021) & 0.096 (0.015) & 0.094	0.014) \\	
                 & lasso        & 0.095 (0.024) & 0.097 (0.022) & 0.097 (0.016) & 0.095	0.014) \\	
                 & m.power  & {\bf 0.077} (0.020) & {\bf 0.091} (0.020) & {\bf 0.094} (0.015) & {\bf 0.093} (0.013) \\	
                 & power      & 0.095 (0.024) & 0.097 (0.022) & 0.098 (0.016) & 0.095	 0.014) \\ \hline
$5\%$     & m.lasso    & 0.210 (0.042) & 0.222 (0.032) & 0.231 (0.027) & 0.231 (0.021) \\	
                 & lasso        & 0.220 (0.042) & 0.225 (0.032) & 0.233 (0.027) & 0.233 (0.021) \\	
                 & m.power  & {\bf 0.193} (0.042) & {\bf 0.216} (0.032) & {\bf 0.226} (0.027) & {\bf 0.228} (0.021) \\	
                 & power      & 0.221 (0.042) & 0.225 (0.032) & 0.233 (0.027) & 0.233 (0.021) \\ \hline	
$10\%$   & m.lasso    & 0.304 (0.043) & 0.333 (0.035) & 0.339 (0.031) & 0.331 (0.028) \\	
                 & lasso        & 0.315 (0.043) & 0.336 (0.035) & 0.342 (0.031) & 0.333 (0.028) \\	
                 & m.power  & {\bf 0.287} (0.043) & {\bf 0.326} (0.035) & {\bf 0.334} (0.031) & {\bf 0.327} (0.028) \\	
                 & power      & 0.315 (0.043) & 0.337 (0.035) & 0.342 (0.031) & 0.333 (0.028) \\ \hline	
$20\%$   & m.lasso    & 0.438 (0.044) & 0.502 (0.047) & 0.510 (0.038) & 0.515 (0.039) \\	
                 & lasso        & 0.449 (0.044) & 0.506 (0.047) & 0.513 (0.038) & 0.517 (0.039) \\	
                 & m.power  & {\bf 0.419} (0.044) & {\bf 0.494} (0.047) & {\bf 0.504} (0.038) & {\bf 0.510} (0.038) \\	
                 & power      & 0.449 (0.044) & 0.507 (0.047) & 0.513 (0.038) & 0.518 (0.039) \\ \hline	
$50\%$   & m.lasso    & 0.695 (0.038) & 0.990 (0.072) & 1.044 (0.076) & 1.056 (0.057) \\	
                 & lasso        & 0.706 (0.038) & 0.997 (0.073) & 1.049 (0.077) & 1.061 (0.058) \\	
                 & m.power  & {\bf 0.676} (0.039) & {\bf 0.979} (0.072) & {\bf 1.034} (0.076) & {\bf 1.048} (0.057) \\	
                 & power      & 0.706 (0.038) & 0.997 (0.073) & 1.049 (0.077) & 1.061 (0.058) \\
\bottomrule
\end{tabular}
\label{tab:bino-estB}
}
\end{table}

\begin{table}[t]
\caption{Prediction of a binomial model. Reported are the mean and standard deviation (in the parenthesis) of the misclassification error for $y$ out of 100 data replications.}
\vspace{0.1in}
\centerline{
\begin{tabular}{clcccc}
\toprule
Sparsity &  & \multicolumn{4}{c}{Rank}  \\ \cmidrule{3-6}
$s$  &  Method  & $R=1$ & $R=5$ & $R=10$ & $R=20$ \\ \midrule
$1\%$     & m.lasso    & {\bf 0.231} (0.029) & {\bf 0.337} (0.030) & {\bf 0.362} (0.024) & {\bf 0.372} (0.024) \\	
                 & lasso        & 0.411 (0.038) & 0.416 (0.037) & 0.413 (0.035) & 0.415 (0.031) \\	
                 & m.power  & 0.252 (0.028) & 0.346 (0.027) & 0.370 (0.025) & 0.382 (0.029) \\	
                 & power      & 0.763 (0.075) & 0.764 (0.071) & 0.777 (0.082) & 0.771 (0.068) \\ \hline	
$5\%$     & m.lasso    & {\bf 0.199} (0.021) & {\bf 0.363} (0.023) & {\bf 0.393} (0.022) & {\bf 0.409} (0.026) \\	
                 & lasso        & 0.459 (0.022) & 0.460 (0.025) & 0.460 (0.030) & 0.461 (0.023) \\	
                 & m.power  & 0.220 (0.025) & 0.368 (0.023) & 0.398 (0.024) & 0.416 (0.025) \\
                 & power      & 0.853 (0.091) & 0.854 (0.088) & 0.824 (0.080) & 0.850 (0.080) \\ \hline
$10\%$   & m.lasso    & {\bf 0.194} (0.023) & {\bf 0.359} (0.027) & {\bf 0.396} (0.024) & {\bf 0.404} (0.027) \\	
                 & lasso        & 0.469 (0.022) & 0.466 (0.022) & 0.467 (0.023) & 0.462 (0.020) \\	
                 & m.power  & 0.217 (0.027) & 0.368 (0.027) & 0.404 (0.024) & 0.413 (0.024) \\	
                 & power      & 0.864 (0.079) & 0.865 (0.095) & 0.860 (0.086) & 0.874 (0.076) \\ \hline
$20\%$   & m.lasso    & {\bf 0.193} (0.030) & {\bf 0.343} (0.027) & {\bf 0.375} (0.027) & {\bf 0.396} (0.026) \\	
                 & lasso        & 0.469 (0.022) & 0.466 (0.025) & 0.470 (0.025) & 0.465 (0.023) \\	
                 & m.power  & 0.217 (0.030) & 0.355 (0.024) & 0.384 (0.027) & 0.403 (0.023) \\	
                 & power      & 0.886 (0.094) & 0.870 (0.091) & 0.859 (0.089) & 0.858 (0.089) \\ \hline
$50\%$   & m.lasso    & {\bf 0.187} (0.023) & {\bf 0.282} (0.033) & {\bf 0.312} (0.032) & {\bf 0.344} (0.027) \\	
                 & lasso        & 0.472 (0.019) & 0.471 (0.025) & 0.470 (0.023) & 0.465 (0.023) \\	
                 & m.power  & 0.214 (0.023) & 0.308 (0.030) & 0.336 (0.028) & 0.359 (0.024) \\	
                 & power      & 0.872 (0.091) & 0.860 (0.089) & 0.869 (0.083) & 0.873 (0.092) \\	
\bottomrule
\end{tabular}
\label{tab:bino-pred}
}
\end{table}

We make the following observations. First, for the normal response, the proposed matrix version of estimators almost always outperform the corresponding vector version in terms of both parameter estimation and prediction, and this holds true for both the lasso and power penalties. The only exception is that when the signal is extremely sparse ($s = 1\%$ of non-zero elements), where the usual vector version of regularized estimators perform slightly better. Second, for the binomial response, the matrix version is superior than the vector version for all ranks and sparsity levels, and such a superiority is more striking in terms of predicting the binary outcome. Third, as for the effects of the signal rank and sparsity, the regular version of regularized estimators perform better when the signal is more sparse (a smaller $s$), while is relatively insensitive to the rank ($R$). By contrast, the proposed matrix version of regularized estimators perform better when the rank is smaller, and is insensitive to the coefficient sparsity. Those patterns agree with our expectations since the former penalizes directly on the coefficient sparsity, whereas the latter penalizes on the rank sparsity. Finally, comparing the lasso penalty with the power penalty, the two yield similar results, while the lasso usually performs better when the rank is large, and the power penalty is better when the rank is small. Such patterns agree with what we observed in Example 1, and can provide some useful guidelines when choosing a penalty function given the data.

\begin{figure}[t]
\centering
\includegraphics[height=2.5in,width=6in]{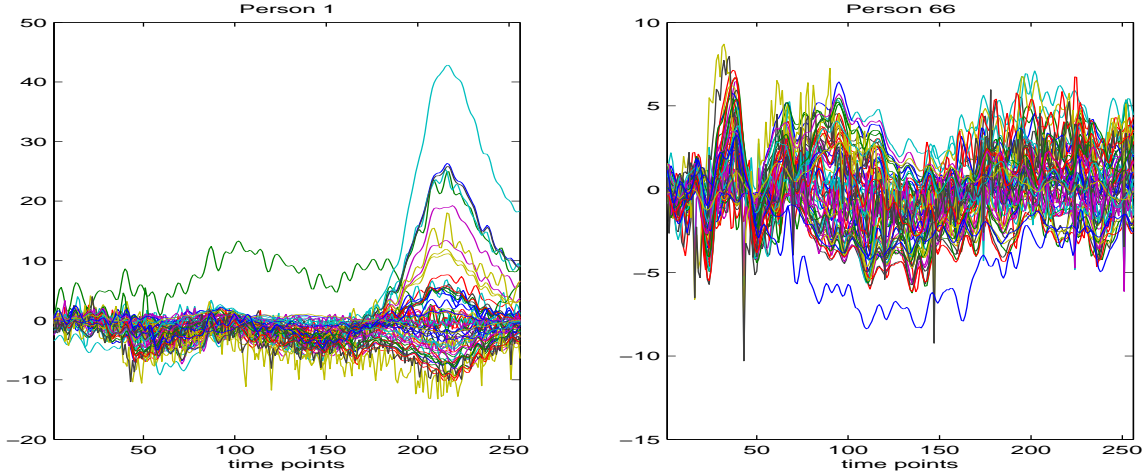}
\caption{EEG signals of an alcoholic subject (left) and a control subject (right). Each curve tracks signals from one electrode.}
\label{fig:EEG-samplepath}
\end{figure}

\medskip
\noindent
\textbf{Example 3: EEG Data Analysis.} We analyzed the motivating EEG data in the Introduction %\citep{FrankAsuncion10UCIMLR}
and compared a number of regularization estimators. The data set consists of 77 individuals with alcoholism and 44 individuals as the control. For each subject, 64 channels of electrodes were placed at different locations of scalp and the voltage values are recorded at 256 time points. Besides, each subject performed 120 trials under three types of stimuli: single stimulus, two matched stimuli and two unmatched stimuli. One primary interest was to study the association between alcoholism and the pattern of voltage values over times and channels. Figure \ref{fig:EEG-samplepath} displays the EEG signals from a randomly chosen alcoholic subject and a control subject, where the horizontal axis is the time, the vertical axis is the voltage, and each curve represents an electrode channel. It is clearly seen that the alcoholic and control subjects demonstrate distinct characteristics.

\citet{LiKimAltman2010} and \citet{HungWang11} both analyzed this same data set. Following their practice, we focused on the data under the single stimulus condition only, and averaged for all the trials under that condition for each subject. The resulting covariates $\xbf_i$ are $256 \times 64$ matrices, and the response $y_i$ is a binary variable indicating whether the $i$th subject is alcoholic ($y_i=1$) or not ($y_i=0$), $i = 1, \ldots, 122$. We applied matrix lasso, lasso, matrix power and power to the data, and evaluated each solution via cross-validation based misclassification error. More specifically, we divided the full data into a training and a testing sample using $k$-fold cross-validation. When $k$ equals the sample size, it is the leave-one-out cross-validation. Then for the training data, we further employed a 5-fold cross-validation to tune the shrinkage parameter $\lambda$. We then applied the tuned model that is fully based on the training data now to the testing data and evaluated the misclassification error rate for testing. Table \ref{tab:EEG} reports the results for leave-one-out, 5, 10, and 20-fold cross-validation results. It is clearly seen that the matrix version of the regularized estimators achieve smaller misclassification errors compared to the vector version counterparts. Comparing the lasso and the power penalties, the lasso achieves a slightly better classification accuracy.

\begin{table}[t]
\caption{Misclassification error rate for EEG data.}
\vspace{0.1in}
\centerline{
\begin{tabular}{lcccc}
\toprule
Method   & leave-one-out & 5-fold & 10-fold & 20-fold \\ \midrule
m.lasso   & \hspace{0.15in} {\bf 0.230} \hspace{0.15in} & \hspace{0.15in} {\bf 0.214} \hspace{0.15in} & \hspace{0.15in} {\bf 0.222} \hspace{0.15in} & \hspace{0.15in} {\bf 0.181} \hspace{0.15in} \\	
lasso       & 0.246 & 0.287 & 0.271 & 0.264\\	
m.power & 0.246 & 0.222 & 0.228 & 0.213\\	
power     & 0.254 & 0.329 & 0.244 & 0.276\\	
\bottomrule
\end{tabular}
\label{tab:EEG}
}
\end{table}

We also make a few remarks regarding to this data analysis. First, \citet{HungWang11} analyzed the same data using a matrix logistic regression, which is a rank-1 special case of the tensor GLM of \citet{ZhouLiZhu12CPTensor}. Their method does not work for $n < p$, and thus a ridge type regularization was employed. The ridge tuning parameter was chosen so that the leave-one-out classification accuracy is maximized, and the corresponding best classification result was reported, which is slightly better than the classification accuracy reported in Table \ref{tab:EEG}. We believe, however, a fair evaluation of prediction should have the parameter tuning solely based on the training data, and then report the corresponding testing error. If one tunes the parameter based on the reported testing error, the result would be over optimistic. Second, both \citet{LiKimAltman2010} and \citet{HungWang11}  preprocessed the EEG data by reducing the dimensionality of the matrices to a smaller scale. Both used principal components analysis (PCA) for reduction, but they employed different variants of PCA for matrices. Part of reason for the dimension reduction preprocessing was that their proposed numerical methods can not directly handle the size of $256 \times 64$ of the EEG data. By contrast, our proposal is not limited by the matrix size and were directly applied to the original data, since our Nesterov algorithm is highly efficient and scalable. On the other hand, we agree that such preprocessing could potentially improve the overall classification accuracy by removing noisy irrelevant information. However, we note that, PCA is known as an unsupervised dimension reduction approach that can not guarantee full preservation of all relevant information. Moreover, it introduces another layer of tuning, including choosing the right version of PCA for matrices, and determining the optimal number of principal components. Our goal of this data analysis has primarily been the comparison of the regularized matrix regression estimators with the vector counterparts, so we have chosen not to include any preprocessing that requires a significant amount of additional tuning.

\section{Discussion}
\label{sec:discussion}

Motivated by modern scientific data arising in areas such as imaging and cytometry, we study in this article the problem of regressions with matrix covariates. Regularization is bound to play a crucial role in such regressions due to the ultrahigh dimensionality and complex structure of the matrix data. We have proposed a class of regularized matrix regression models by penalizing the spectrum of the matrix parameters. It is based upon the observation that the matrix signals are often of, or can be well approximated by, a low rank structure. Consequently, the new method focuses on the sparsity in terms of the rank of the matrix parameters rather than the number of nonzero entries, and is intrinsically different from the classical lasso and related penalization approaches.

In many applications, sparsity is sought in certain pre-specified \emph{basis} systems rather than the original coordinates. Specifically, the systematic component takes the form
\begin{eqnarray*}
\eta = \gammabf \trans \Zbf + \langle \Bbf, \Sbf_1 \trans \Xbf \Sbf_2 \rangle,
\end{eqnarray*}
where $\Sbf_j \in \real{p_j \times q_j}$ contains $q_j$ basis vectors for $j=1,2$. For 2D images, over-complete wavelet basis ($q_j > p_j$) are often used in both dimensions. For the EEG data that are channels by time points, wavelet or Fourier basis can be applied to the time dimension. It is of direct interest to seek a sparse low rank representation of the signal $\Bbf$ in the basis system by solving the regularized regression. In that sense, our proposed regularized matrix regression can be considered as the matrix analog of the classical basis pursuit problem \citep{ChenDonohoSaunders01BasisPursuit}.

We have concentrated on problems with matrix covariates throughout this article. In applications such as anatomical magnetic resonance imaging (MRI) and functional magnetic resonance imaging (fMRI), the covariates are in the form of multidimensional arrays, a.k.a.\ tensors ($D>2$). It is natural to extend the work here to regularized tensor regressions. However, the problem formulation requires an appropriate norm for tensors that is in analogous to the nuclear norm for matrices, and the involved regularization and optimization are to be fundamentally different from the methods for matrices. We are currently pursuing this line of extension, and will report the results elsewhere.

\baselineskip=15pt

%% for use by Hua
%\bibliography{../../../bib-HZ}
%\bibliographystyle{apalike}

% for use by Lexin
\bibliography{matrix_lasso_notes}
\bibliographystyle{apalike}

\end{document}